\documentclass[pdflatex,sn-mathphys-num]{sn-jnl}

\usepackage{graphicx}%
\usepackage{multirow}%
\usepackage{amsmath,amssymb,amsfonts}%
\usepackage{amsthm}%
\usepackage{mathrsfs}%
\usepackage[title]{appendix}%
\usepackage{xcolor}%
\usepackage{textcomp}%
\usepackage{manyfoot}%
\usepackage{booktabs}%
\usepackage{algorithm}%
\usepackage{algorithmicx}%
\usepackage{algpseudocode}%
\usepackage{listings}%
\usepackage{subcaption}

\theoremstyle{thmstyleone}%
\theoremstyle{thmstyletwo}%

\theoremstyle{thmstylethree}%

\begin{document}

\title[Fractal spatio-temporal scale-free messaging: amplitude modulation of self-executable carriers given by the Weierstrass function's components]{\vspace{-2.5cm}Fractal spatio-temporal scale-free messaging: amplitude modulation of self-executable carriers given by the Weierstrass function's components}


\author*[1,2,3]{\fnm{Hector} \sur{Zenil}}\email{hector.zenil@kcl.ac.uk}
\equalcont{These authors contributed equally to this work.}

\author[3]{\fnm{Luan Carlos de Sena Monteiro} \sur{Ozelim}}\email{luan.ozelim@algocyte.ai}
\equalcont{These authors contributed equally to this work.}

\affil[1]{\orgdiv{School of Biomedical Engineering \& Imaging Sciences}, \orgname{King’s College London}, \orgaddress{ \city{London}, \postcode{WC2R 2LS}, \country{United Kingdom}}}


\affil[2]{\orgname{The Alan Turing Institute}, \orgaddress{\street{British Library}, \city{London}, \postcode{NW1 2DB}, \country{United Kingdom}}}

\affil[3]{\orgname{Oxford Immune Algorithmics}, \orgaddress{\street{Davidson House, The Forbury}, \city{Reading}, \postcode{RG1 3EU}, \country{United Kingdom}}}


\abstract{In many contexts of communication, the communication capabilities of the involved actors cannot be known beforehand, whether it is a cell, a plant, an insect, or even a life form unknown to Earth. Regardless of the recipient, the message space and time scale could be too fast, too slow, too large, or too small and may never be decoded. Therefore, it pays to devise a way to encode messages agnostic of space and time scales. We propose the use of fractal functions as self-executable infinite-frequency carriers for sending messages given their properties of structural self-similarity and scale invariance. We call it `fractal messaging'. Starting from a spatial embedding, we introduce a framework for a space-time scale-free messaging approach to this challenge. When considering a space and time-agnostic framework for message transmission, it would be interesting to encode a message such that it could be decoded at several spatio-temporal scales. Hence the core idea of the framework proposed herein is to encode a binary message as waves along infinitely many frequencies (in power-like distributions) and amplitudes, transmit such a message, and then decode and reproduce it. To do so, the components/cycles of the Weierstrass function, a known fractal, are used as carriers of the message. Each component will have its amplitude modulated to embed the binary stream, allowing for a space-time-agnostic approach to messaging.}




\maketitle

\section{Introduction}\label{sec1}

Despite the multitude of formal definitions, the word `communicate' is historically related to the word `common'. Etymologically, the word `communicate' can be traced back to the Latin verb \textit{communicare} which means `to share', `to make common', which itself is derived from the Latin root \textit{communis} that translates as common. Therefore, there is general agreement that when someone or something communicates, one makes things common \cite{Rosengren}. There are no pre-established criteria that parties to a given act of communication must meet,
so that communication may take place between entities of very different complexities except perhaps some elements of a sharing reality.

So-called models of communication are used as conceptual representations, which aim to provide a simplified overview of the main components involved in the process of communication~\cite{Ruben}. Normally, it is assumed that two entities, the sender and the receiver, exchange something, a signal, that could be meaningful to both. It is true, on the other hand, that the sender and the receiver do not need to be individuals or singular entities; they may be groups. Whatever the case, it is often understood that both the sender and the receiver are known. 

To make things as common as possible when it comes to unknown entities or unknown properties of entities, it is important to study communication as agnostically as possible based on universal first principles. In the context of making things common, it is crucial for the communication process that both the sender and the receiver can find some meaning in the signal. Besides, the signal itself must be perceivable by both these entities. Regarding the question of meaning, whenever the sender and receiver are known, some general correspondence between the signal and its meaning can be guaranteed. On the other hand, when only the sender is known, to try to establish any comprehension, the means used to transmit the signal and the signal itself should be the simplest possible, using basic elements such as electromagnetic waves, as noted in pioneering works on interstellar communications \cite{COCCONI1959}. 

Binary signals have been used to send messages to space in the expectation that they could be decoded and understood by receivers in other parts of the universe. One such example is the Arecibo message sent to space in the 1970s. Recently, an updated, binary-coded message has been developed~\cite{Jiang2022}. Such messages may include basic mathematical and physical concepts, enabling the establishment of a universal means of communication. They may contain information on the biochemical composition of life on Earth, the Solar System's time-stamped position in the Milky Way relative to known globular clusters, as well as digitised depictions of the Solar System and of Earth's surface. More recently, we published work on a method based on the principles of (algorithmic) information theory to agnostically decode properties of these messages in an optimal and universal fashion, allowing a naive receiver to infer the geometric properties of the message, a first step in any attempt at understanding~\cite{etpaper}. In another piece of work, we connected the presence of life and possibly intelligence to the same principles of (algorithmic) information, with life serving 
as empirical evidence of low randomness and high structure in the universe~\cite{zenil2012}.


While our previous work covered a specific challenge on the receiving end, the sender has to make multiple assumptions and also faces some challenges. For a signal to be perceived by an entity--before it can begin to attempt to decode or interpret it--the signal should be within its dimensional spatio-temporal scales. One challenge being that of different spatio-temporal scales across entities. Even on Earth, it appears to us as animals that cells or plants are simple or static with little to no communication capabilities because their time scales differ significantly by several orders of magnitude from that of humans.

There has been considerable work done on the exchange of information between various organisms, most of which can be compiled under the field of biocommunication \cite{Seckbach}. During interspecies communications, it may be challenging to precisely characterise the means used to transmit a given message, since different species have particular biostructures which enable (or disable) the adequate flow of signals. For example, plants have rigid cell walls that restrict movement and usually prevent them from sending and receiving signals that depend on rapid movement \cite{Karban,Schenk2010}.

Movement is deeply related to space and time, such that communication may be prevented if both sender and receiver function at different spatio-temporal scales. Regardless of the recipient's nature, the message space and time scale could be too fast, too slow, too large or too small and may well never be decoded. Therefore, it would pay to devise a way to encode messages agnostic of time scale. Fractal messaging would allow one to do this. 

We propose that fractals represent an opportunity as self-exectuable carriers for messages which need to be scale-independent, especially the ones with structural self-similarity or even scale invariance. Starting from a spatial embedding rationale, a framework will be developed for a time scale-free messaging alternative. Thus, in the present paper, a theoretical framework will be developed to provide time-agnostic communication, herein understood as the transmission of binary messages. A viable suggestion for message embedding and transmission based on software and hardware implementations will be discussed, as well as how the framework enables time scale-free messaging.

\section{Preliminaries}


\subsection{Fractals}

According to Mandelbrot, a fractal is a set whose Hausdorff-Besicovitch dimension strictly exceeds its topological dimension \cite{Mandelbrot1982}. A remarkable characteristic of fractals is that they can assume a similar geometrical form
at various scales, a feature
referred to as self-similarity. If this form is exactly the same at every scale, the set is called a self-affine or a scale-invariant set.
\cite{Falconer_1985,gouyet1996physics}. As discussed in the Introduction, fractals are interesting candidates for carriers of signals, lending to these their self-similarities. 

\subsubsection{The Mandelbrot set}

The Mandelbrot set is a set of complex numbers (therefore two-dimensional if plotted in the complex plane) defined by a simple recursion formula that hides intricate patterns and great complexity. It can be defined as the set of complex numbers $c$ for which the function $f_c(z)=z^2+c$ remains finite after being infinitely iterated from $z=0$. This set was first defined and drawn by Brooks and Matelski \cite{Brooks} in 1978, but became famous in 1980 when Mandelbrot \cite{Mandelbrot1980} obtained high-quality visualisations of it.

When plotted in the complex plane, the boundary of the Mandelbrot set produces infinitely complicated patterns, especially when zoomed in. Such a boundary is a fractal curve, following Mandelbrot`s definition. Despite its simple definition, the fact that the set is bidimensional may pose challenges to the message embedding process. Thus, lower-dimensional alternatives such as the Weierstrass function should be considered.

\subsubsection{The Weierstrass function}

\begin{figure}[!ht]
\centering
  \includegraphics[width=0.9\textwidth]{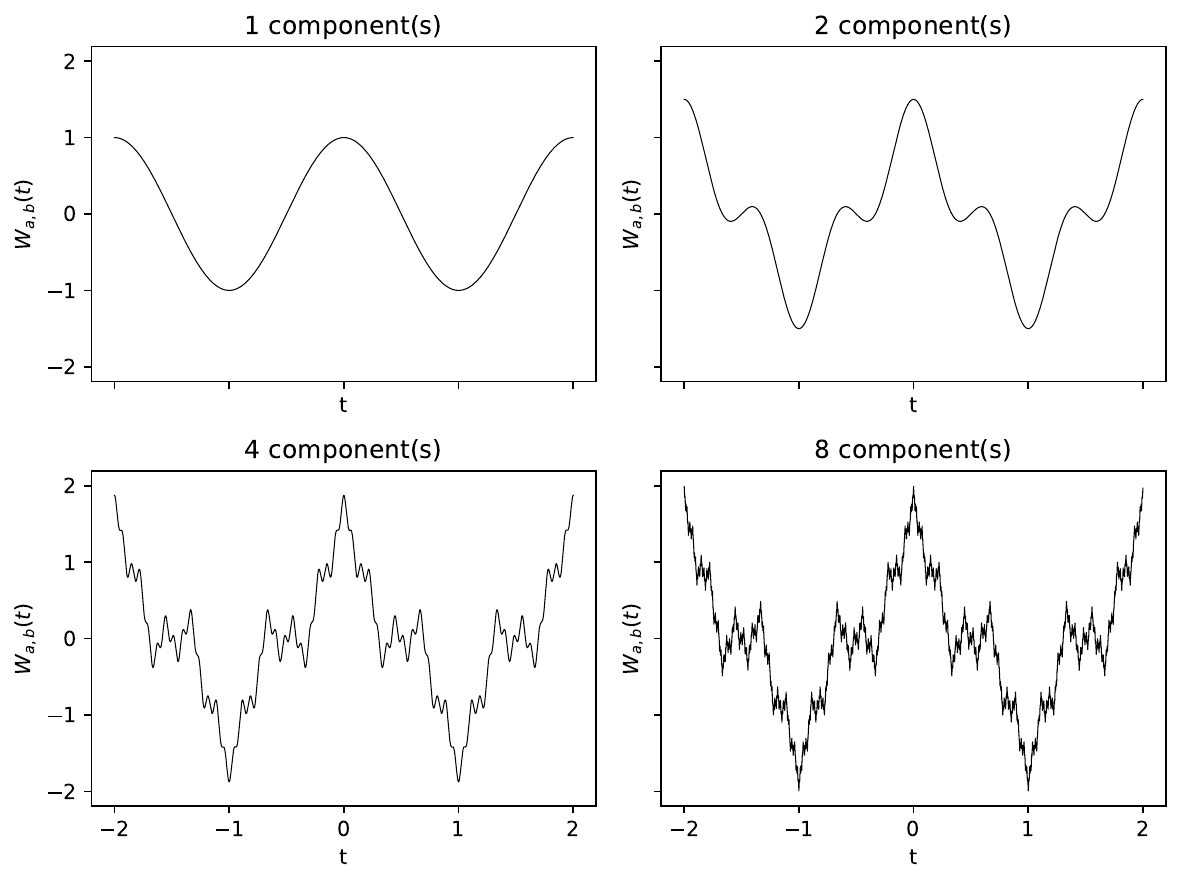}
  \caption{General behaviour of $W_{a,b}(t)$ as more terms (components) are considered in the summation when $a=1/2$ and $b=3$.}  
  \label{mth1}
\end{figure}

Weierstrass defined a function which is continuous but nowhere differentiable as the following Fourier series:

\begin{equation}\label{weir}
W_{a,b}(t)=\sum_{n=0} ^\infty a^n \cos(b^n \pi t),
\end{equation}

\noindent where $0<a<1$, $b$ is a positive odd integer, $t \in \mathbb{R}$ and

\begin{equation}
    ab > 1+\frac{3}{2} \pi.
\end{equation}

Hardy \cite{hardy} further discussed these restrictions, showing that whenever $0<a<1<b$ and $ab \geq 1$, for $b$ not necessarily an integer, the function does not possess a finite differential coefficient at any point. This latter set of restrictions will be considered in the present paper.

Figure \ref{mth1} presents the general behaviour of $W_{a,b}(t)$ as the number of terms (or components) in the summation increase. The figure indicates that $W_{a,b}(t)$ is nothing but the superposition of sinusoidal waves, which have been extensively used as carriers for messages using traditional radio transmission techniques. In the next section, a specific technique for amplitude modulation is discussed.

\subsection{Amplitude modulation of signals}

There are several ways to encode binary messages using carrier signals. In the present paper, the so-called Amplitude Shift Keying (ASK) scheme will be considered, where one simply multiplies the wave form of a certain carrier wave by another signal formed by the sum of several rectangular waves representing the bits.

\begin{figure}[!ht]
\centering
  \includegraphics[width=0.9\textwidth]{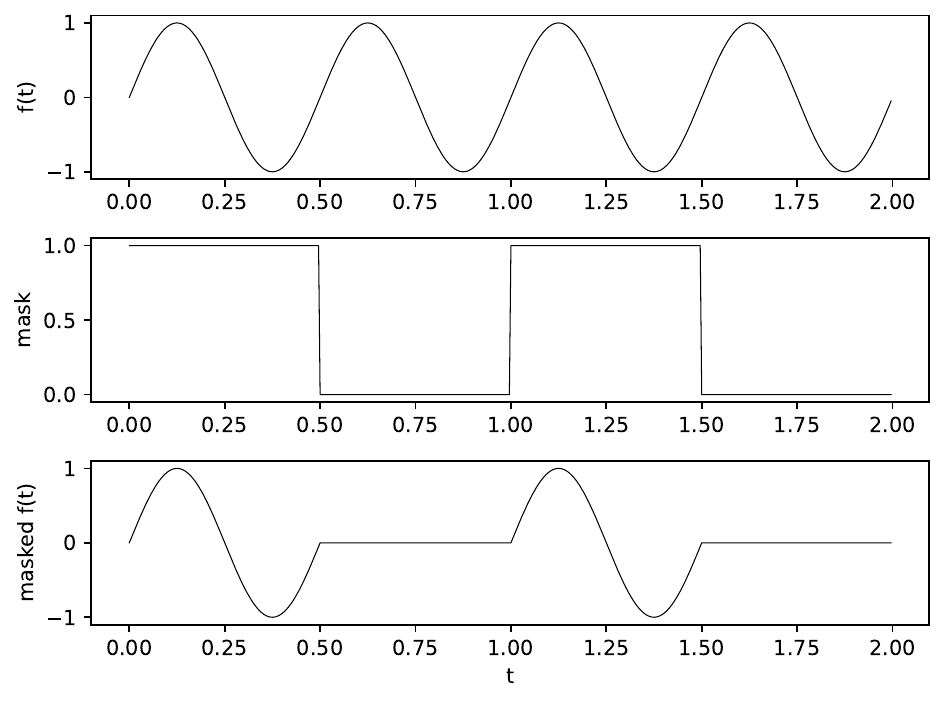}
  \caption{Simple illustration of amplitude modulation (or masking) of a signal $f(t)$.}  
  \label{mth2}
\end{figure}

Figure \ref{mth2} presents a simple modulation of the signal $f(t)$. Considering that each bit modulates a single period of the signal $f(t)$, it can be seen that the binary ``10'' was used to modulate it.

Now that all the preliminary concepts have been presented, the general intuition behind fractal messaging can be discussed.

\section{Intuition}

\begin{figure}
\centering
\begin{subfigure}{0.49\textwidth}
    \includegraphics[width=\textwidth]{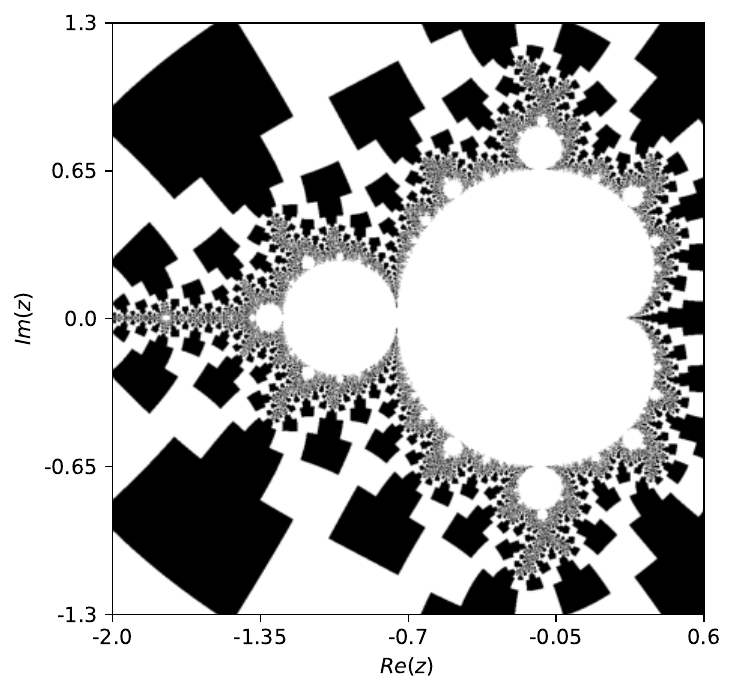} 
\end{subfigure}
\hfill
\begin{subfigure}{0.49\textwidth}
     \includegraphics[width=\textwidth]{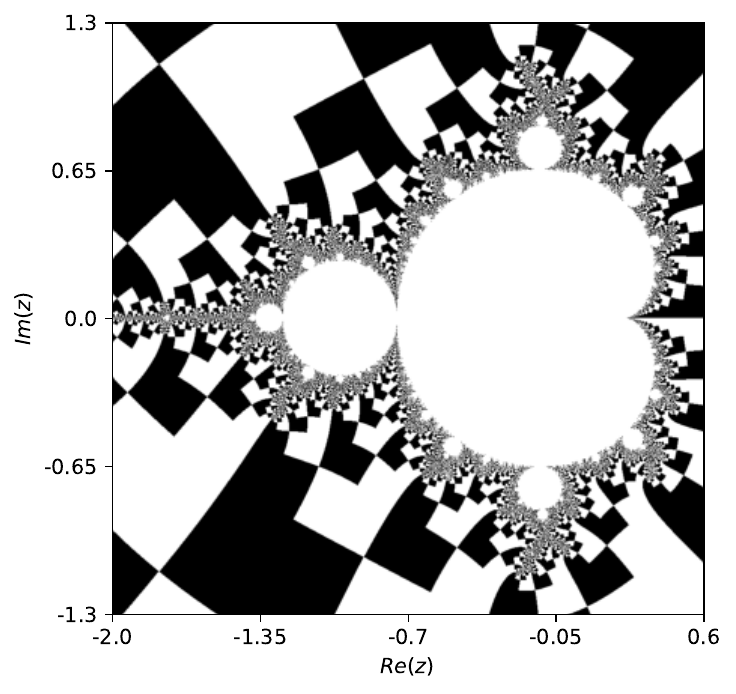} 
\end{subfigure}
\caption{Simple illustration of embedding $Re(z_e) \geq 0$ (left) and $Im(z_e) \geq 0$ (right) for $R_e = 10$, that is, the encoding of the message `01' in the Mandelbrot set. Other fractals can be used for the same purposes, including, for example, the Koch snowflake, which has a mathematical description between the Mandelbrot 2D nature and a discrete version of the signal structure of the Weiertrass function.}
\label{MandReIM}
\end{figure}

Formally, one can define the recurrence relation used to build the Mandelbrot set as follows: Let $z_0 = 0$ and $z_{n+1} = z_n^2 + c$, for a fixed complex number $c$. The points belonging to the set are the values of $c$ such that $\lim\limits_{n\to \infty} |z_n| < R_e$, where $R_e$ is the escape radius chosen for the algorithm. It is easy to prove using the triangle inequality that if $|z_n|>2$, then the sequence will diverge, implying that $R_e=2$ is a natural choice, although other values of $R_e \geq 2$ can be considered.

For the values not belonging to the Mandelbrot set, let $z_e$ represent the minimum value of $z_{n}$ such that $|z_{n-1}| < R_e$ and $|z_{n}| > R_e$,  where $|w|$ is the absolute value of the complex number $w = w_{R} + i w_{I}$. In that case, $w_{R} = Re(w)$ and $w_{I} = Im(w)$ are the Real and Imaginary parts of $w$, respectively. Then, since the values of $z_e$ are unique for a given escape radius, it is possible to use them to perform color encodings of information into the complement of the Mandelbrot set in the complex plane. For example, for binary encodings, Figure \ref{MandReIM} presents the plots obtained for $Re(z_e) \geq 0$ and $Im(z_e) \geq 0$, respectively. Both figures illustrate that one can embed messages, such as a binary counter representing all positive integers or
natural numbers, in the complement of the Mandelbrot set, respecting the same properties of self-similarity, and therefore invariance, of the sender/receiver spatial scales.

Note that when considering the message to be transmitted as a bidimensional image, some sense of orientation is mandatory in order for the recipient to properly understand it, as in the case of the Arecibo message, which becomes an unintelligible set of characters if not arranged properly. This can be challenging to ensure in practice. Besides, the reproduction of images could rely on specific sensory characteristics (vision, for example), which are not guaranteed to be among the receiver entity's capacities. This has been explored by Vakoch \cite{Vakoch} while addressing the challenge of communicating information unique to one civilisation, underscoring the fact that sending two-dimensional pictorial representations of mathematical concepts and physical objects may not be readily intelligible to a civilisation with different conventions than ours. 

On the other hand, Vakoch \cite{Vakoch} indicated that the use of three-dimensional representations redundantly encoded in multiple formats (e.g., as both vectors and as rasters) could be better suited for the communication task. Overall, one of the core ideas presented is that, after a gradual process of presenting familiar objects (stars, for example), unfamiliar concepts could be introduced and that redundancy could increase the likelihood of intelligibility. 

The general idea of redundancy as a means to attain intelligibility is deeply related to the use of fractals as carriers for messages, but from a transmission point of view rather than in light of the content of the message itself. In this context, devising a way to transmit messages such that redundancy is achieved both spatially and temporally is of interest, especially if this could be implemented in a compact unidimensional spatial setting. Such an approach will be described in the next section.

\section{Using the Weierstrass function's components as carriers for binary messages}

When considering a space and time-agnostic framework for message transmission, it would be interesting to encode a message so that it could be decoded at several space and time scales. One would assume that any ``being'' receiving the message would live at a specific time scale, and that messages could be ``understood'' as long as they were decoded and reproduced at suitable frequencies, for example.

Hence the core idea of the framework proposed herein is to encode a binary message along infinitely many frequencies (with power-like distributions), transmit such a message and then decode and reproduce it. If one were to start with uni-dimensional messages, which are simply streams of binary information, a natural carrier would be uni-dimensional electromagnetic waves. Thus, it is of interest to consider fractals which could be viewed as uni-dimensional electromagnetic waves. This is the case with the Weierstrass function and its components.

Figure \ref{mth2} indicates how to modulate a single sinusoidal signal in order to embed a binary stream in it. For different types of receiver entities, each functioning on a given spatio-temporal scale, the message could only be perceived if reproduced as a suitable signal in terms of amplitude and frequency. Therefore, since the correct scales are unknown, a direct way to guess them is by repeating the message across the whole frequency-amplitude spectrum. In order to cover infinitely many frequency ranges, one can modulate each of the components of the summation in Eq. \ref{weir}. In order to ensure a single-period single-bit modulation rationale, one must make the rectangular waves representing each bit match a single period of the carrier wave. This can be done by adjusting the frequencies and performing phase shifts on the rectangular waves. Mathematically, let $A_m(t,msg)$ be the $m$-th amplitude modulation function representing the same binary message $msg$. The resulting fractal communication signal $\mathcal{F}_{a,b}$ can be represented as:

\begin{equation}\label{fractm}
    \mathcal{F}_{a,b}(msg,t)=\sum_{m=0} ^\infty a^m A_m(t,msg) \cos(b^m \pi t),
\end{equation}

It is possible to obtain $A_m(t,msg)$ by the following three-step procedure:

\begin{itemize}
    \item Generate a single pulse for each bit in the binary stream $msg$;
    \item Multiply each pulse by the value of each bit in the stream;
    \item Sum all the pulses and repeat them indefinitely.
\end{itemize}

Thus, by following the procedure described, it is now possible to define:

\begin{eqnarray}
    A_m(t,msg) = \sum_{i=1}^{L} msg_i \left[ u\left(mod\left(t-\frac{2(i-1)}{b^m},\frac{2L}{b^m}\right)\right) \right. - \nonumber \\ \left. u\left(mod\left(t-\frac{2(i-1)}{b^m},\frac{2L}{b^m}\right) -\frac{2}{b^m}\right) \right]
\end{eqnarray}

\noindent where $L$ is the number of bits in $msg$, $msg_i$ is the $i$-th bit in $msg$, $mod(x,y)$ is the remainder of the division of $x$ by $y$ and $u(.)$ is the Heaviside step function, hereby defined as being 0 for negative arguments and 1, otherwise.

The series representing $\mathcal{F}_{a,b}$ is convergent whenever the restrictions for $a$ and $b$ presented for $W_{a,b}$ are satisfied, since:

\begin{equation}
    \mathcal{F}_{a,b}(msg,t) \leq W_{a,b}(t).
\end{equation}

By considering equation \ref{fractm} as the signal to be transmitted, not only infinitely many frequencies are covered (controlled by $b$), but also infinitely many amplitudes (controlled by $a$).

\subsection{Creating the fractal communication signal}

In order to make full use of equation (\ref{fractm}), the following pseudo-code can be used:

\begin{algorithm}
\caption{Generate $\mathcal{F}_{a,b}(msg,t)$}\label{algo1}
\begin{algorithmic}[1]
\Require $a$, $b$, $M$ (number of components of $W_{a,b}(t)$ to be considered); $T$ (duration of the signal); $msg$ (message to be transmitted).
\State $msg_{binary} \Leftarrow $ a binary encoding for $msg$ (the ASCII binary representation of strings in $msg$, for example).
\For{$m \in \{0,1,...M\}$}
        \State $C_m(t) \Leftarrow $ the $m$-th component of $W_{a,b}(t)$, which is a sinusoidal wave of amplitude $a^m$ and frequency $b^m/2$.
        \State $mask_m(t) \Leftarrow $ a mask composed of square waves juxtaposed (phase shifted) to represent each each bit in $msg_{binary}$. Each rectangular wave must have the same frequency as $C_m$.
        \State $masked_{m}(t) \Leftarrow C_m \times mask_m$, which represents the masked component.
\EndFor
\State $\mathcal{F}_{a,b}(msg,t) \Leftarrow \sum_{m=0}^M masked_{m}(t)$.
\State Transmit $\mathcal{F}_{a,b}(msg,t)$ from $t=0$ to $t=T$.
\end{algorithmic}
\end{algorithm}

\begin{figure}[!ht]
\centering
  \includegraphics[width=0.9\textwidth]{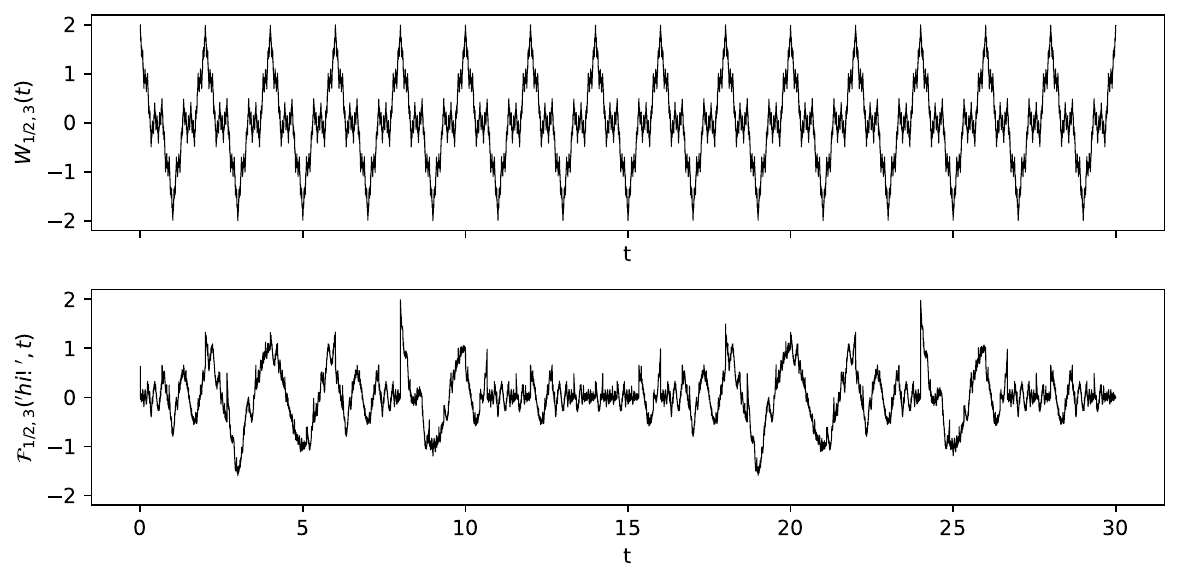}
  \caption{Encoded message ``hi!'' using the components of $W_{1/2,3}$ as carriers. Ten components were considered, for computational simplicity.}  
  \label{mth3}
\end{figure}

As a simple example, one can encode the message ``hi!'' using its ASCII binary representation `011010000110100100100001'. Considering positive time as our $t$ variables, Figure \ref{mth3} presents the encoded fractal message.

\begin{figure}
\centering
\begin{subfigure}{0.9\textwidth}
    \includegraphics[width=\textwidth]{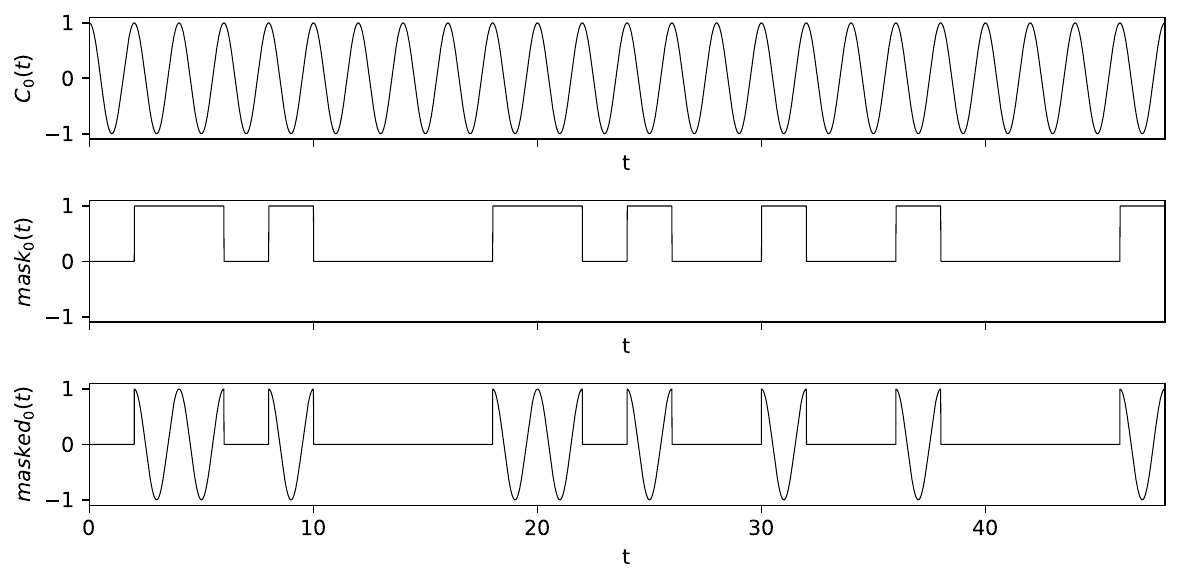} 
\end{subfigure}
\hfill
\begin{subfigure}{0.9\textwidth}
     \includegraphics[width=\textwidth]{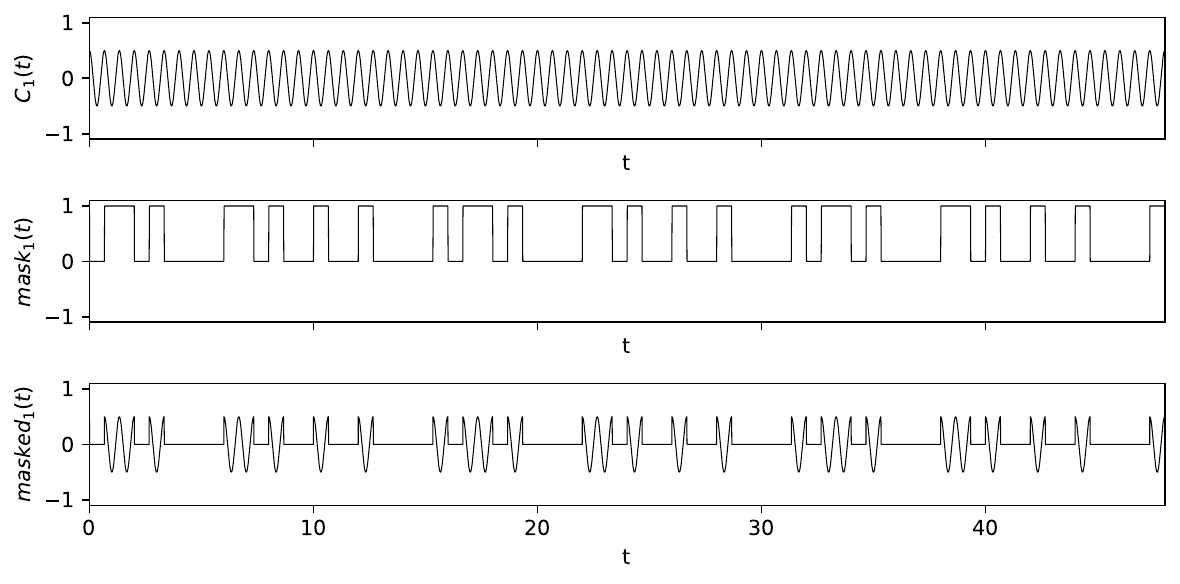} 
\end{subfigure}
\hfill
\begin{subfigure}{0.9\textwidth}
     \includegraphics[width=\textwidth]{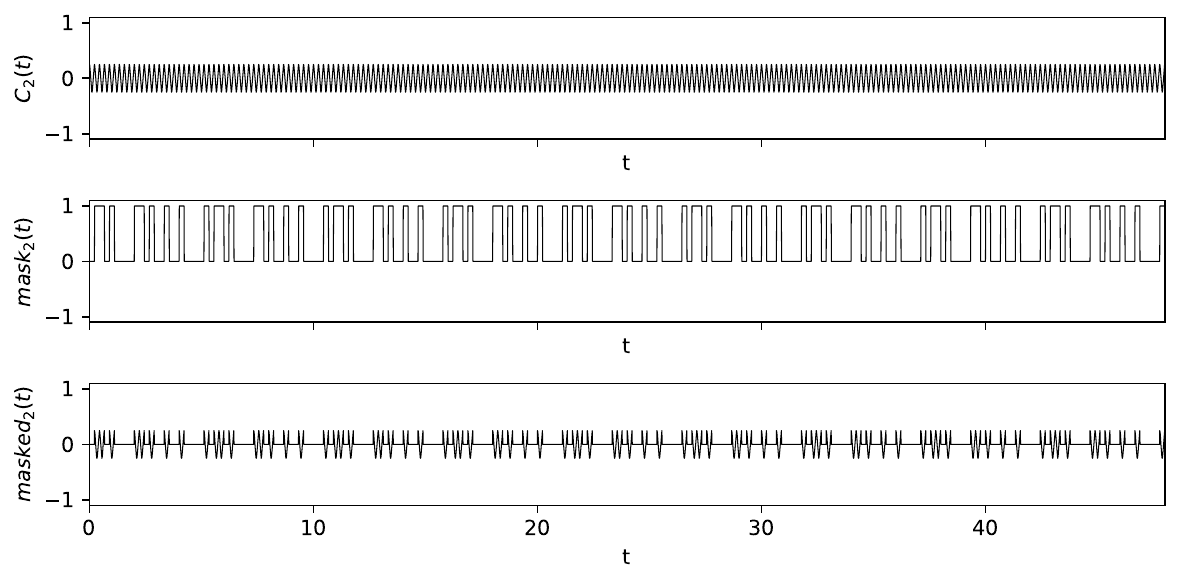} 
\end{subfigure}
\caption{First three components of the fractal message embedding the ASCII representation of  ``hi!''}
\label{comp0}
\end{figure}

In order to create the fractal message in Figure \ref{mth3}, Algorithm \ref{algo1} was implemented in Python. Thus, to illustrate each step in the \textbf{for} loop, let one consider Figure \ref{comp0}.

\begin{figure}[!ht]
\centering
  \includegraphics[width=0.9\textwidth]{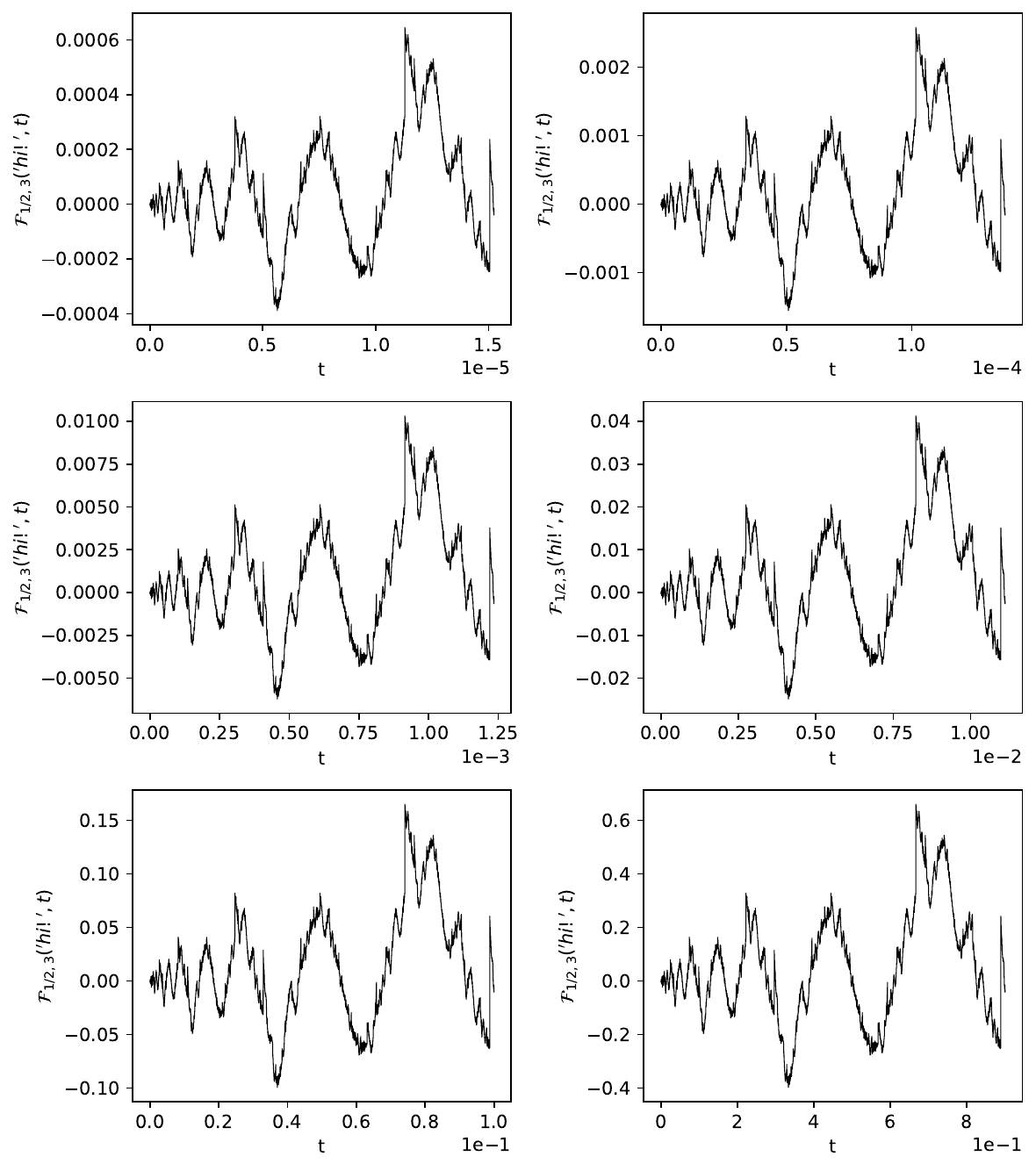}
  \caption{Zooming in $\mathcal{F}_{1/2,3}('hi!',t)$. Twenty components were considered, for computational simplicity. They are all the same but at different scales, encoding all the same message `hi' an infinite number of times.}  
  \label{mth4}
\end{figure}

Figure \ref{comp0} presents the first three iterations of the \textbf{for} loop in Algorithm \ref{algo1}. For $C_0$, the $0$-th component of the Weierstrass function, a simple sinusoidal wave of amplitude $(1/2)^0$ and frequency $3^0/2$ is seen. Then, the ASCII binary stream `011010000110100100100001' is simply represented as a series of square waves side by side (one for each bit in the stream). This sequence of square waves is then multiplied by each element of the carrier (time-wise multiplication), leading to $masked_0$. For the $1$-th component, a similar process is carried out. Now, the amplitude of the sinusoidal wave is $(1/2)^1$ while its frequency is $3^1/2$. It can be seen that the square waves now have both their phase shifting and frequencies adjusted in order to match the properties of the $1$-th component. The same time-wise multiplication of elements is carried out to obtain $masked_1$. This process is also considered for the $2$-th component, with even smaller amplitudes and greater frequencies. After collecting $M$ components (defined according to the numerical precision sought), the complete fractal message can be obtained by summing all the $masked_m$, for $m \in \{0,1,...,M \}$. To produce the lower plot in Figure \ref{mth3}, 10 components were considered. Fractal behaviour is observed for the encoded message by zooming in the lower plot in Figure \ref{mth3}. These zoomed in regions are presented in Figure \ref{mth4}.

It is interesting to note that after embedding the message, the fractal dimension of the resulting signal changes. In the case of Figure \ref{mth3},  it was believed that $W_{a,b}$ had a fractal dimension of $2 + \log_{b}(a)$ \cite{Falconer_1985}, such a result being later proved for some values of $a$ and $b$ \cite{Shen2017}. Thus, $W_{1/2,3}$ has a fractal dimension of $2 - \log_{3}(2) \approx 1.36907$, which could be verified using a Box-counting algorithm. On the other hand, using the same algorithm, the fractal dimension of $\mathcal{F}_{1/2,3}('hi!',t)$ was found to be approximately $1.42146$, slightly higher than the fractal dimension of $W_{1/2,3}$. This suggests that by embedding messages, even though the ASK scheme used actually suppresses information (by not transmitting the signal in the positions of the bit $0$ in the message), the overall complexity of the signal increases as the fractal dimension can be considered its proxy \cite{Mandelbrot1982}. 

\subsection{Decoding and reproducing}

It is possible to use inverse Fourier transform tools to retrieve the original message because the amplitude modulation behaviour repeats itself periodically (infinite loop). In order to reproduce the message, either sounds (pressure waves) or laser pulses could be generated. One caveat when using these approaches is the need to fine-tune the energy content of the reproduction. This is important to avoid under- or over-detection (the laser, for example, could be undetectable in a high energy - high frequency scenario or could be as powerful as the solar system's sun in a low energy - low frequency scenario).

Another interesting approach is to use one-dimensional chains of trapped ions to produce a quantum mechanical system with discrete scale invariance and fractal-like time dependence \cite{Lee2019}. In such a case, it was shown that patterns similar to the Weierstrass function could be produced, which are pivotal to encode the messages using the framework proposed herein. In light of the findings of Rose and Wright \cite{Rose2004}, sending such chains of trapped ions together with circuits to modulate the amplitude of the waves emitted could be an interesting transmission approach at hardware level. There are certainly limitations, but in principle, for a certain range of amplitudes and frequencies, transmission and emission could be achieved. Regarding software implementations, there are also limits up to which the waves can be generated and reproduced. These arise from shortfalls in the precision of the calculations considered, for example.

\subsection{Beyond binary encodings of characters: Iterated Function Systems (IFSs) and Turing machines as fractals}

In the present paper, the messages being transmitted are binary streams representing, for example, ASCII encodings of known characters. Using fractal carriers, the message and the carriers became a single entity, which was shown to be a fractal. This suggests a more general framework where concepts could be stated in terms of fractals, which could ultimately encode not only characters but also computational concepts, in an integrated fractal way. 

As a matter of fact, using fractals to establish a communication system may be beneficial as the universe has been shown to present fractal-like structures with remarkable self-similarity (the self-similar multifractal distributions of dark matter and gas, for example \cite{Gaite2010}). Therefore, the entities receiving the message could be familiar with self-similar structures, due to their occurrence in their own environment. Not only that, but the literature indicates that digital sundials have been devised as projections of shadows of a fractal set \cite{Falconer1987}, which would allow the possible communication of a correspondence between time scales related to Earth's sun and the receiver's sun. Therefore, using fractals as communication entities could be of interest.

Regarding possible relations between fractal geometry and computation, Penrose has suggested that fractals might be a graphical way of looking at non-recursive mathematics \cite{Penrose}. Further exploring this rationale, Dube \cite{dube2} established a relationship between the classical theory of computation and fractal geometry. In particular, he considered the concepts behind IFSs to define fractals, indicating that for every Turing machine (in the context of computation over integers and rationals) there exists a fractal set which can be viewed, in a certain sense, as geometrically encoding the complement of the language accepted by the machine. This suggests that in a fractal-based communication system, it would be possible to transmit fractal-based geometrical models of computation which are computationally universal. Inspired by previous works \cite{Falconer1987}, Dube \cite{dube2,dube} proposed a fractal-based ``sun-computer", indicating that for every Turing machine $M$ there exists such a fractal set $A$ which graphically ``encodes" the behaviour of $M$ on all inputs. This could be a potential carrier for a Turing Machine computational model without the need to use binary encodings.

\section{Conclusion}\label{sec13}

We have introduced a new mathematical framework based on fractal theory that allows spatio-temporal scale-independent and parallel infinite-frequency communication. 

Our approach instantiates the idea of sending a self-encoding/self-decoding signal as a mathematical formula that carries both the data and its unfolding instructions. This would be similar to sending a self-executable message with a computer on which to run it. It is equivalent to sending self-executable computer code that, when unfolded, reads a message at all possible time scales and in all possible channels simultaneously. Other fractal functions may be used and yet to explore is whether they offer any advantage given that the Weierstrass function seems to solve the main identified challenges. However, other fractals in higher dimensions may offer alternative embeddings for multi-modal non-linear messages.

Another aspect to explore is how and whether we are prepared to receive and detect a signal with these properties. Together with our previous work~\cite{etpaper}, we have contributed to specific technical challenges at the receiver and now sender ends, and combining both approaches is to be done.

\backmatter

\bibliography{sn-bibliography}

\newpage

\section*{Supplementary information}

\subsection*{Alternative understanding of the fractal communication process}

\begin{figure}[h]
\centering
\includegraphics[width=0.3\textwidth]{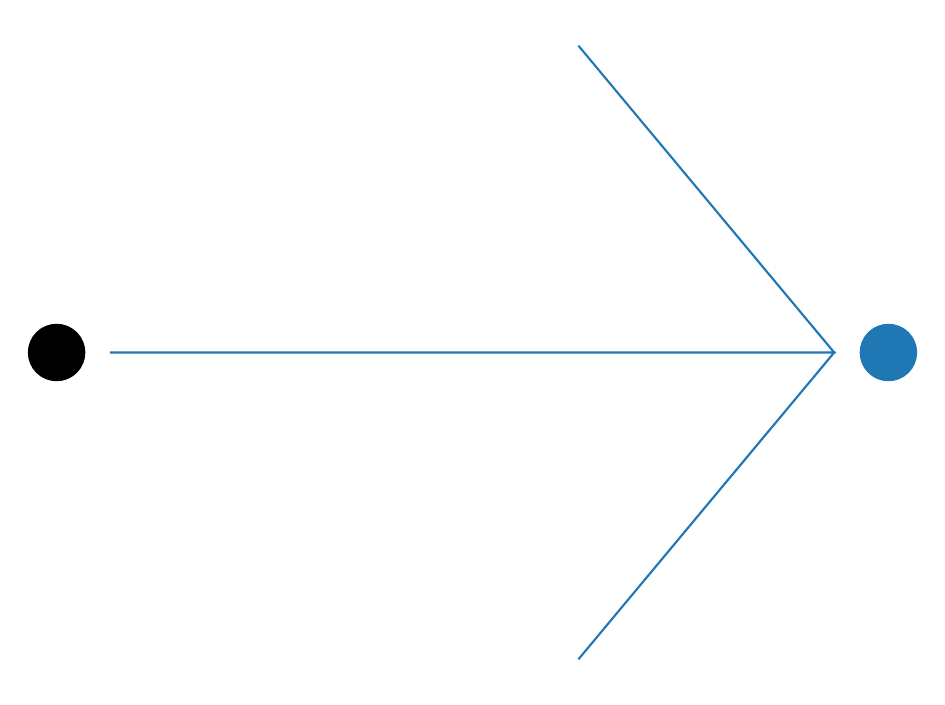}
\caption{\label{fig:0}General communication between sender (black circle) and receiver (blue circle) using a carrier represented as the blue arrow.}
\end{figure}

In general, the communication process encompasses two entities: the sender and the receiver. These entities exchange a message which is transmitted by means of a carrier (or channel) that must be adequate to both contain the message and be perceived by the receiver. Figure \ref{fig:0} represents this process, where the sender is identified as a black circle, the receiver by a blue circle and the message in its carrier is represented as a blue arrow. It is worth noticing that the arrow is colored the same as the receiver because the carrier used must be ``understandable" by the receiver, both in temporal and spatial scales. The width of the arrow can be understood as the spatial scale of the carrier used (and, therefore, of the receiver). 

\begin{figure}[h]
\centering
\includegraphics[width=0.3\textwidth]{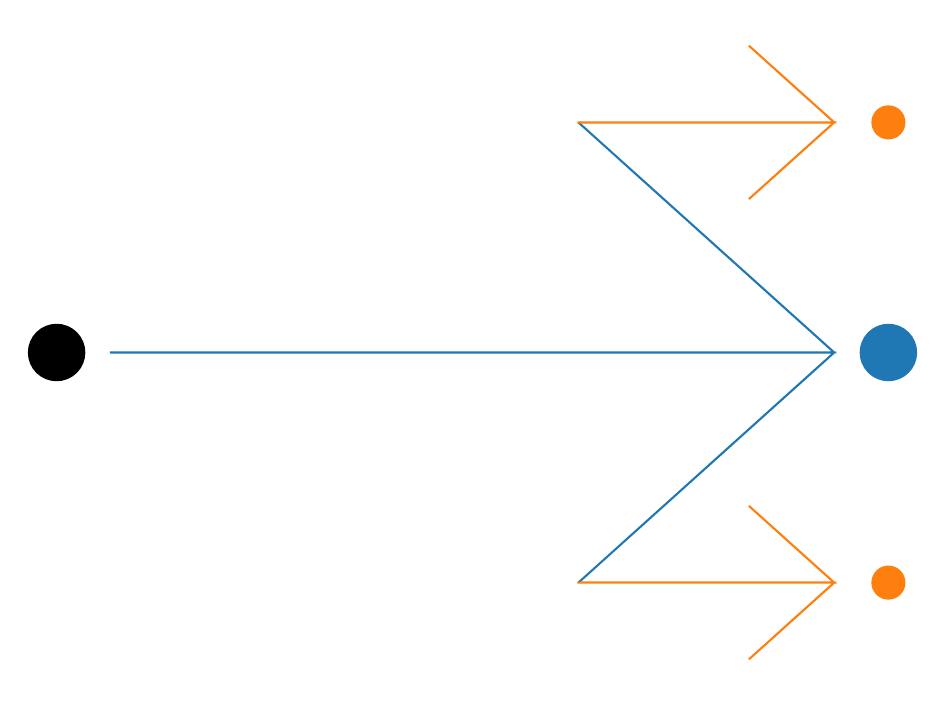}
\caption{\label{fig:1}Communication process with added carrier to make the messsage reach new receivers depicted as orange circles.}
\end{figure}

Since we don't know what/who the receiver will be, we must send the message through carriers which span several spatial-temporal scales. This would ensure that, regardless of the receiver, the message could be received. The closest we can get to covering all the spatial-temporal scales is to actually cover infinitely many of them. This can be achieved by incorporating more and more carriers (channels) to our transmission process. Let one consider Figure \ref{fig:1}. In such figure, another channel is added to our overall carrier, allowing receivers represented as orange circles to properly receive the message.

\begin{figure}[h]
\centering
\includegraphics[width=0.7\textwidth]{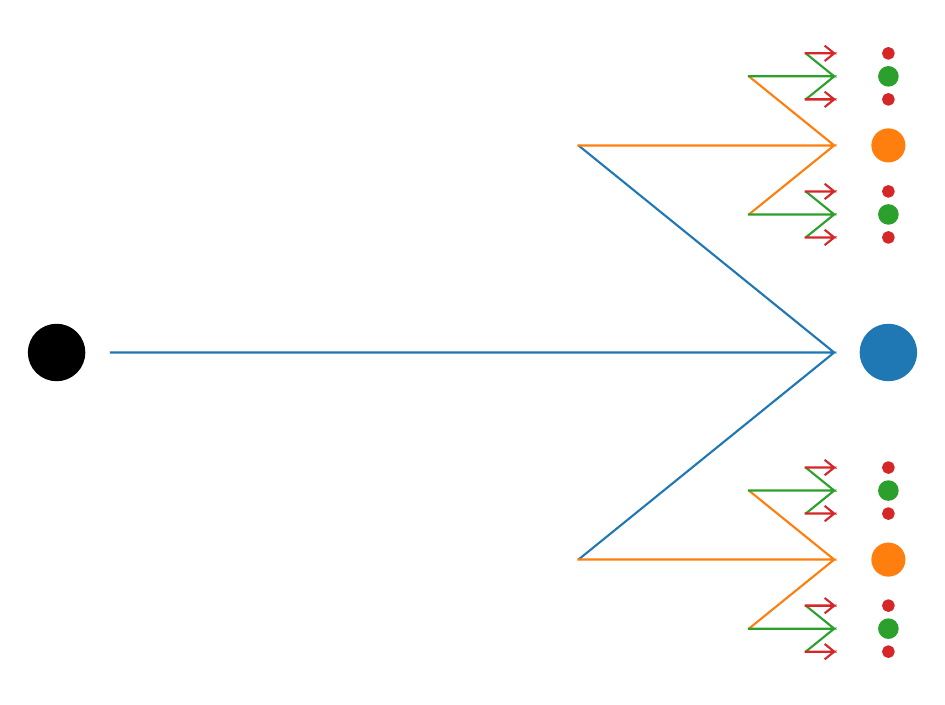}
\caption{\label{fig:2}Communication process with added carriers to make the messsage reach new receivers depicted as green and red circles.}
\end{figure}

It can be seen that the new orange channel allows for receivers in smaller spatial scales to receive the message (smaller arrow widths). Besides that, it can be seen that for the orange channel, two repetitions of the message are transmitted during the same period that the single blue arrow transmits a message. Thus, not only smaller spatial scales are seen, but also higher frequency transmissions. The same process of adding new channels with decreasing spatial scales and increasing frequency can be repeated infinitely. For the next two iterations, Figure \ref{fig:2} represents how receivers represented as green and red circles are reached by the message.

This process of iteratively increasing channels to to build a carrier is exactly the creation of a fractal. These structures are mathematical entities which repeat themselves at different scales and, despite their high complexity, are generated by repeating a simple process (adding more arrows with different scales, as indicated). Thus, a fractal carrier, which encompasses channels along infinitely many spatial-temporal scales, is the foundation of the fractal communication process devised in the paper. Even though Figure \ref{fig:2} is a schematic representation of the fractal communication scheme, the arrow generated by combining all the generated ones is a fractal arrow.

\subsection*{Implementation of the fractal communication}

In the paper, each of the channels represented as arrows in Figures \ref{fig:0} to \ref{fig:2} were considered as a sinusoidal waves, which are standard carriers for long distance communications. In order to use each of these channels to transmit a message, we must embed the message in them. Messages can be any type of information, but since any data can be converted to binary streams by suitable procedures, for convenience they were considered as such. For example, text can be converted to binary streams by using their ASCII binary representations.

\begin{figure}
\centering
\includegraphics[width=0.95\textwidth]{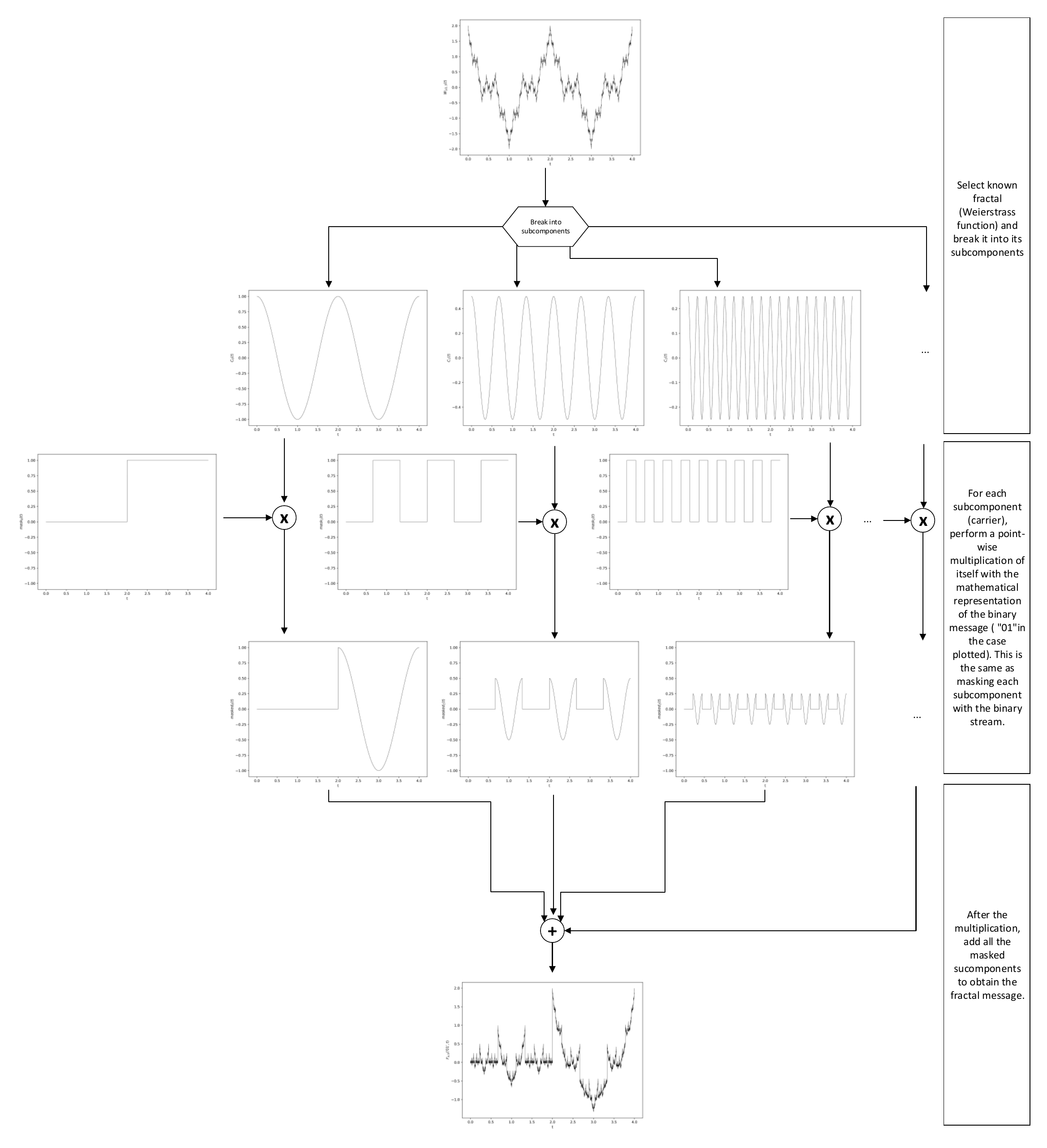}
\caption{\label{fig:flow}Graphical representation of the fractal message creation.}
\end{figure}

The general steps considered in the paper are represented graphically in Figure \ref{fig:flow}. In short, by considering a known fractal (the Weierstrass function), we decompose it into its sinusoidal subcomponents, hereby named carriers or channels. Then, for each of these carriers, we embed a binary stream representing the message we want to send (which is a masking process). Finally, by summing all the masked subcomponents, the fractal message is created.

In order to embed a binary stream into a sinusoidal carrier, we used a simple amplitude modulation scheme, which changes the amplitudes of certain parts of the wave according to the message which is being transmitted. Figure \ref{fig:3} presents this process, which can be described as:

\begin{figure}[h]
\centering
\includegraphics[width=0.8\textwidth]{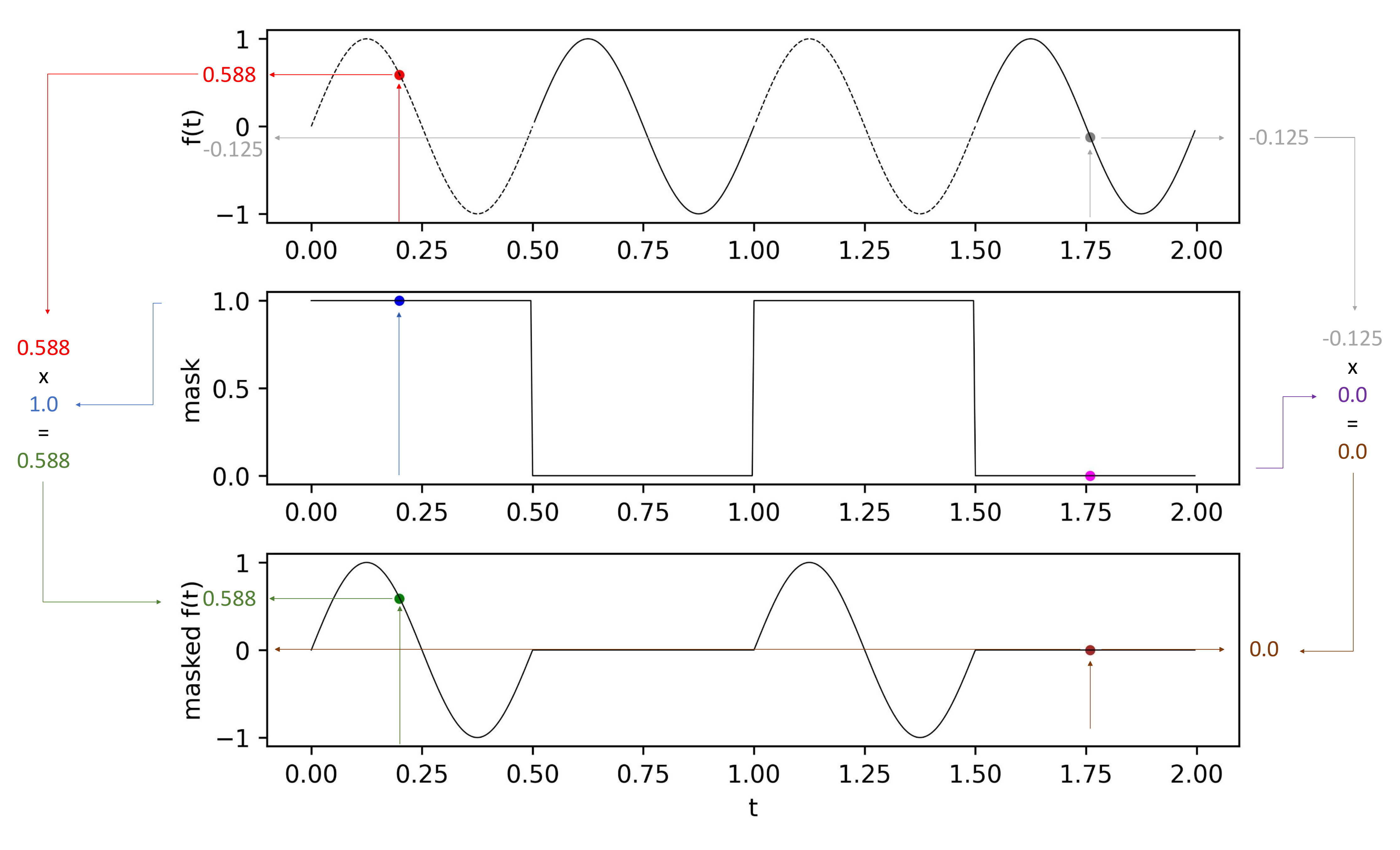}
\caption{\label{fig:3}Embedding a binary stream ``10" into a sinusoidal carrier.}
\end{figure}

\begin{figure}[h]
\centering
\includegraphics[width=0.9\textwidth]{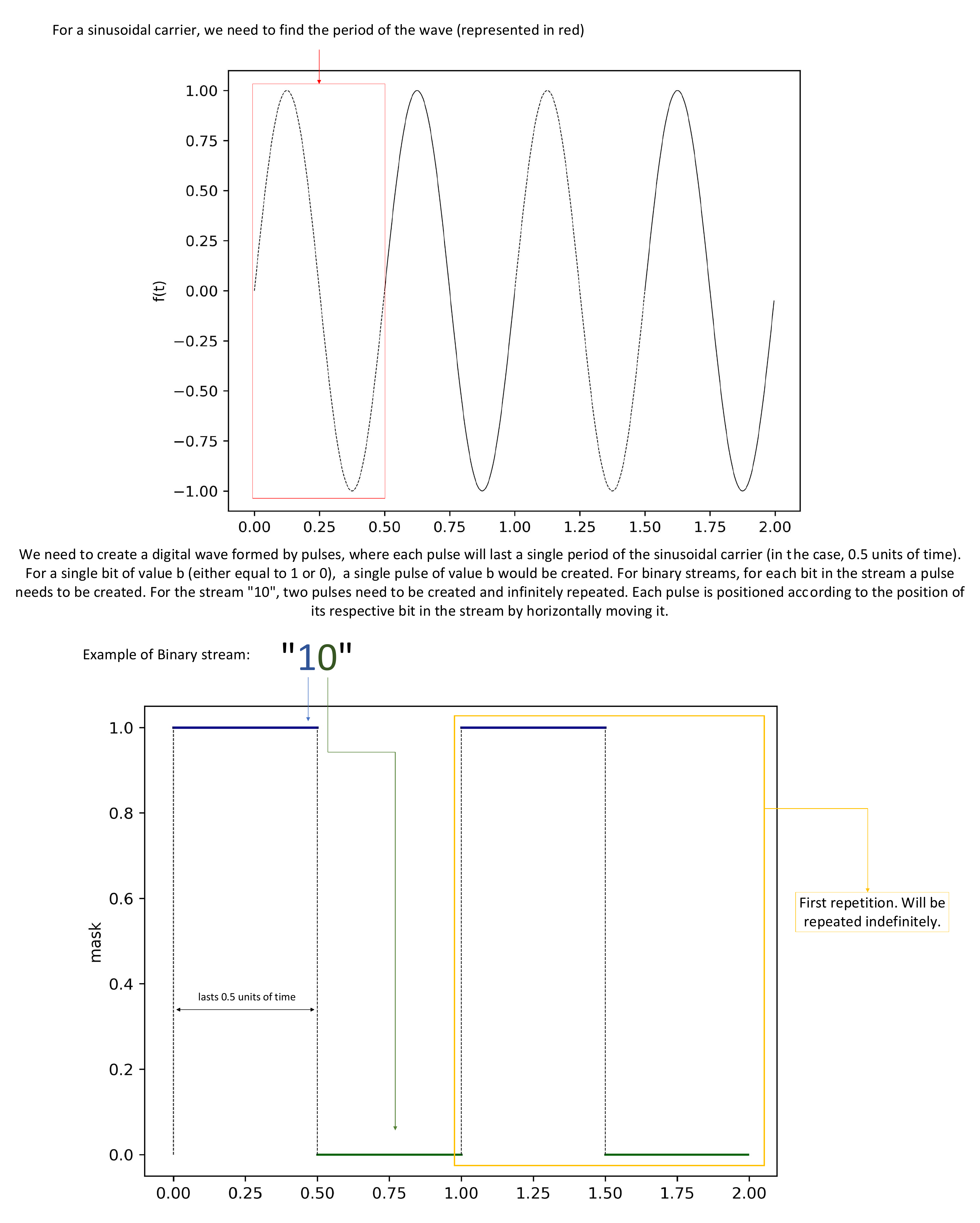}
\caption{\label{fig:44}Creating the mathematical representation of the stream ``10".}
\end{figure}

\begin{itemize}
    \item Consider a sinusoidal wave represented as $f(t)$ (first plot in Figure \ref{fig:3}). We want to embed the binary stream ``10" in it.
    \item Then, for every point in $f(t)$, say at $t=t_1$, we will get the value of $f(t_1)$ and multiply by the mathematical representation of the binary stream ``10" (the second plot in Figure \ref{fig:3}), obtaining the embedded message as the third plot in Figure \ref{fig:3}.
    \item In order to do the multiplication, we must make sure that the period of the sinusoidal wave is the same as the mathematical representation of the binary stream ``10". This can be achieved by making sure that the duration of each pulse in the binary stream is the same as the dashed and full parts of the first plot in Figure \ref{fig:3}. Figure \ref{fig:44} illustrates how the mathematical representation of a binary stream is created.
    \item So, for example, let one take the red point in the first plot in Figure \ref{fig:3}. We can see that the corresponding point in the binary stream is the blue one (same $t$-value, i.e., horizontal coordinate equal to $0.2$). Thus, we get the value of the vertical coordinate in the first plot for the red point ($0.588$) and multiply it by the value of the vertical coordinate of the blue point in the second plot ($1.0$), obtaining the value of the the vertical coordinate of the green point ($0.588$) in the last plot of Figure \ref{fig:3}. This process is repeated for every point in the two upper plots, generating the third plot. Same applies, for example, for the multiplication of the vertical coordinates of the gray (first plot) and magenta (second plot) points, generating the coordinates brown point (third plot). 
    \item By repeating this process for infinitely many arrows (sinusoidal waves), we can add all these signals to obtain the fractal message.
\end{itemize}

If the message to be transmitted is an infinite stream of 1s, the fractal message will be simply the Weierstrass function, a known fractal. By embedding different binary streams, different fractals are obtained.

\subsection*{Computational entities as fractals}

Let us revisit sundials. It is known that these ``clocks" work by exploiting the fact that when the Earth rotates on its axis, the Sun moves across the sky, causing objects to cast shadows. During this process, whenever the sun changes its position, so does the shadow cast by a given reference element. By carefully matching shadow positions with the corresponding hours and physically engraving this matching process, the sun dial becomes a conversion machine whose inputs are light rays and outputs, hours and minutes. A similar process can be conceived for more general operations than the one described.

\begin{figure}[h]
\centering
\includegraphics[width=0.9\textwidth]{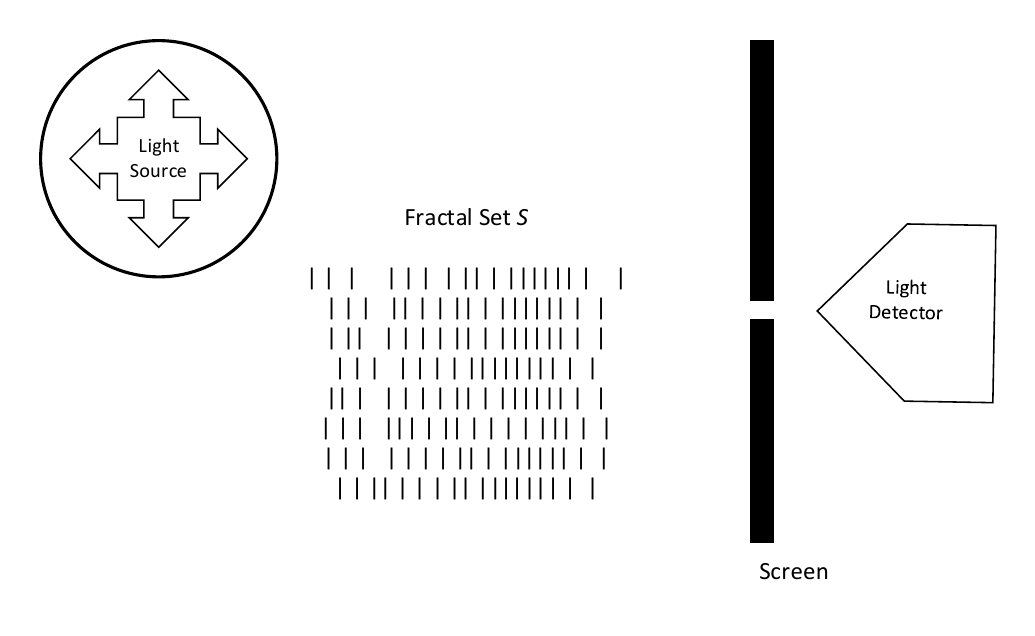}
\caption{\label{fig:45}Schematic of a fractal-based ``sun-computer".}
\end{figure}

Previous researchers have shown that the so-called Iterated Function Systems (IFSs) are computational entities deeply linked to fractals. This opens a wide range of possibilities for fractal communication of not only text and data, but also of computational entities themselves. Simant Dube proposed a fractal-based ``sun-computer", depicted in Figure \ref{fig:45}, indicating that for every Turing machine $M$ there exists such a fractal set $S$ which graphically ``encodes" the behaviour of $M$ on all inputs. Thus, to observe the output of the Turing Machine for an input $w$, by changing the positions of the light source and of the screen, the light receptor would reveal the output of the Turing machine for $w$ without the need of any other calculations. This could be a potential carrier for a Turing Machine computational model without the need to use binary encodings.

\end{document}